\RequirePackage{fix-cm}
\documentclass[twocolumn]{svjour3}          % twocolumn
\smartqed  % flush right qed marks, e.g. at end of proof
\usepackage{graphicx}
\usepackage{amssymb}
\usepackage{amsmath}
\begin{document}

\title{Protonated Clusters of Neon and Krypton%\thanks{Grants or other notes
}


\author{Michael Gatchell         \and
        Paul Martini \and Arne Schiller \and Paul Scheier %etc.
}


\institute{M. Gatchell, P. Martini, A. Schiller, P. Scheier \at
              Institut f\"{u}r Ionenphysik und Angewandte Physik, Universit\"{a}t Innsbruck, Technikerstr.~25, A-6020 Innsbruck, Austria \\
             % Tel.: +123-45-678910\\
              \email{michael.gatchell@uibk.ac.at}           %  \\
           \and
           M. Gatchell \at
              Department of Physics, Stockholm University, 106 91 Stockholm, Sweden
}

\date{Received: date / Accepted: date}

\maketitle

\begin{abstract}
We present a study of cationic and protonated clusters of neon and krypton. Recent studies using argon have shown that protonated rare gas clusters can have very different magic sizes than pure, cationic clusters. Here we find that neon behaves similarly to argon, but that the cationic krypton is more similar to its protonated counterparts than the lighter rare gases are, sharing many of the same magic numbers.

\end{abstract}

\section{Introduction}
\label{intro}
Clusters of rare gas atoms are elegant systems for studying the packing of spherically symmetric particles into highly symmetric geometries. They were the subject of much interest in the mass spectrometry community from the 1980s on, when measurements of charged clusters formed in supersonic expansions showed the characteristic magic numbers associated with icosahedral (sub-)shells \cite{Echt:1981aa,Ding:1983aa,Harris:1984aa,Scheier:1987aa,Levinger:1988aa,Miehle:1989ab}. More recently it was shown that small impurities can dramatically change the magic numbers of charged rare gas clusters. A high resolution mass spectrometric study of cationic and protonated argon clusters showed that only the latter exhibited the characteristic magic numbers associated with icosahedral structures \cite{Gatchell:2018aa}. For the pure clusters, the presence of a compact Ar$_3^+$ charge center distorts the structures of the clusters, preventing the efficient packing of atoms in icosahedral shells. Theoretical calculations showed that for the protonated systems, the charge-carrying proton forms a bridge between two Ar atoms, nearly preserving the Ar-Ar separation and the overall symmetry of the neutral systems \cite{Gatchell:2018aa,Giju:2002aa}. This reduces the strain on the structures, giving the Ar$_n$H$^+$ clusters the same magic numbers as for model Lennard Jones systems and neutral Ar$_n$ \cite{Gatchell:2018aa,Wales:1997aa}. These findings helped explain the discrepancies seen between different studies of charged argon clusters, a topic that had been discussed for over 30 years \cite{Ding:1983aa,Harris:1984aa,Scheier:1987aa,Levinger:1988aa,Miehle:1989ab,Ferreira-da-Silva:2009aa,Harris:1986aa,Milne:1967aa}. 

Here we expand upon the experimental work of protonated rare gas clusters by studying pure cationic and protonated clusters of neon and krypton that are produced in doped superfluid helium nanodroplets. Neon and krypton surround argon in the periodic table and by comparing the new results with the previous findings for Ar, we are able to identify trends in the behavior of charged rare gas clusters and their protonated counterparts. 
\begin{figure*}
  \includegraphics[width=6.5in]{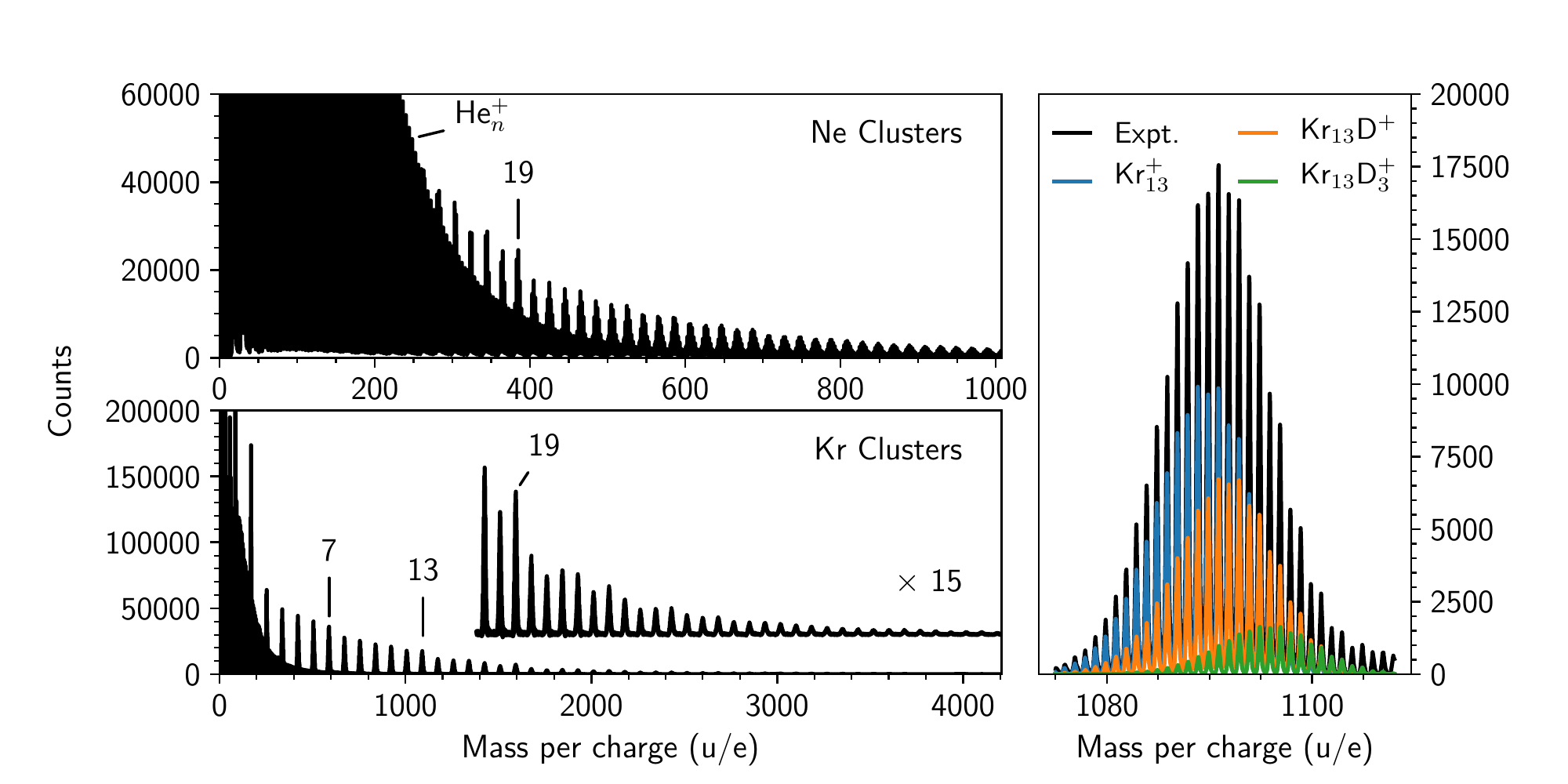}
\caption{Mass spectra for positively charged products after helium nanodroplets doped with Ne and H$_2$ (top left)} or Kr and D$_2$ (bottom left) are ionized by 78\,eV electrons. Cluster series for the pure and protonated rare gases are clearly visible up to sizes of 50 atoms. The right panel is a zoom-in of the Kr$_{13}^+$ region showing the large number of individual peaks that arise from the isotopic pattern of Kr. Here we also show the fitted distributions of the main complexes in this mass window. The Kr$_{n}$D$_3^+$ series (potentially also containing contributions from Kr$_{n}$HeD$^+$) is included to better constrain the contribution from Kr$_{n}$D$^+$ but is beyond the focus of this work.
\label{fig:MS}       % Give a unique label
\end{figure*}

\section{Methods}
\label{sec:meth}

\paragraph{Experimental Setup} 
Rare gas clusters are produced in superfluid helium nanodroplets using the setup described in detail in Refs.\ \cite{Schobel:2011aa,Kurzthaler:2016aa,Kuhn:2016aa}. Droplets of He containing on average a few million atoms are formed by the expansion of compressed (2.1\,MPa) He gas through a 5\,$\mu$m nozzle that is cooled to 8.3\,K. The droplets pass though a 0.8\,mm skimmer positioned 8\,mm downstream from the nozzle before entering a pair of pickup chambers. Here the droplets capture Ne or Kr and H$_2$/D$_2$ that are introduced though gas inlets, rare gases in the first chamber and hydrogen/deuterium in the second, which condense into clusters in the superfluid 0.37\,K droplets. Deuterium is used for the Kr clusters due to their complex isotopic pattern as the higher mass will increase the separation of peaks in the mass spectra. The droplets are ionized by impact of 78\,eV electrons which produces He$^{+}$ near the surface of the droplets. The charge will then typically migrate via resonant hole-hopping through the droplet before forming a He$_2^{+}$  \cite{Scheidemann:1993aa,Ellis:2007aa,Mauracher:2018aa}. This ion will then move though the droplet, attracted by the higher polarizability of the dopant clusters compared to the surrounding He, ionizing the dopant by electron transfer in a highly exothermic process. The charged dopants are then often expelled from the droplets, giving bare clusters. The positively charged products are analyzed using a reflectron time-of-flight mass spectrometer (Tofwerk AG model HTOF) with a rated $m/\Delta m$ resolution of 5000. The mass spectra are calibrated and analyzed using the IsotopeFit software \cite{Ralser:2015aa}, which efficiently deconvolutes overlapping peaks in the mass spectra and corrects for isotopic distributions. The method of producing rare gas clusters from helium nanodroplets has in the past been used to study, He$_n^+$ \cite{Schobel:2011aa}, Ar$_n^+$ \cite{Ferreira-da-Silva:2009aa,Gatchell:2018aa} and Kr$_n^+$ \cite{Schobel:2011aa} clusters, giving results in good agreement with other techniques \cite{Echt:1981aa,Ding:1983aa,Harris:1984aa,Scheier:1987aa,Levinger:1988aa,Miehle:1989ab,Milne:1967aa}. 

\paragraph{Theoretical Tools}

We have used the Gaussian 16 software \cite{Frisch:2016aa} to perform electronic structure calculations of neutral, cationic, and protonated Ne and Kr clusters. Cluster geometries were optimized at the MP2(Full)/def2-SVPP level of theory and a vibrational frequency analysis was performed on each optimized structure to ensure that potential energy minima are obtained.

\section{Results and Discussion}
\label{sec:res}

Mass spectra of Ne and Kr clusters born in helium nano\-droplets are shown in Figure \ref{fig:MS}. In both cases there is a clear contribution from He$_n^+$ clusters with $n$ up to about 200 that are rest products of the droplets from the ionization process. In the analysis of the Ne and Kr cluster series, the contributions from He-containing peaks are corrected for. The Ne and Kr cluster series are clearly visible and extend up to at least 50 atoms. Anomalies are visible in both cases, with 19 being the clearest magic number. For Kr, which has less overlap with the He series, abundance anomalies are also visible at $n=7$ and 13. Both measurements contain mixtures of the pure rare gas clusters and clusters that also contain hydrogen, mainly in the form of a single proton/deuteron that is formed by the breakup of H$_2$/D$_2$ during ionization. From the overview spectrum alone it is not clear magic numbers arise from the pure rare gas clusters and which ones come from the protonated counterparts. Furthermore, even in experiments studying pure clusters, small amounts of residual water can effectively contribute with protons when the doped droplets are ionized. Careful analysis is thus required to determine the magic numbers associated with each species.

%Both of these measurements were performed without introducing hydrogen gas into the second pickup chamber and should represent the series of pure Ne$_n^+$ and Kr$_n^+$ clusters. However, even very small amounts of residual water can effectively contribute with protons the doped droplets are ionized. Carefully analyzing the mass spectra of the pure clusters shows that there certainly is a contribution from Ne$_n$H$^+$ and Kr$_n$H$^+$ clusters, and that this contribution increases with cluster size, consistent with the scaling of pickup probability with increasing He droplet size (giving both larger rare gas clusters and greater likelihood of picking up residual water). 

\begin{figure}
  \includegraphics[width=3in]{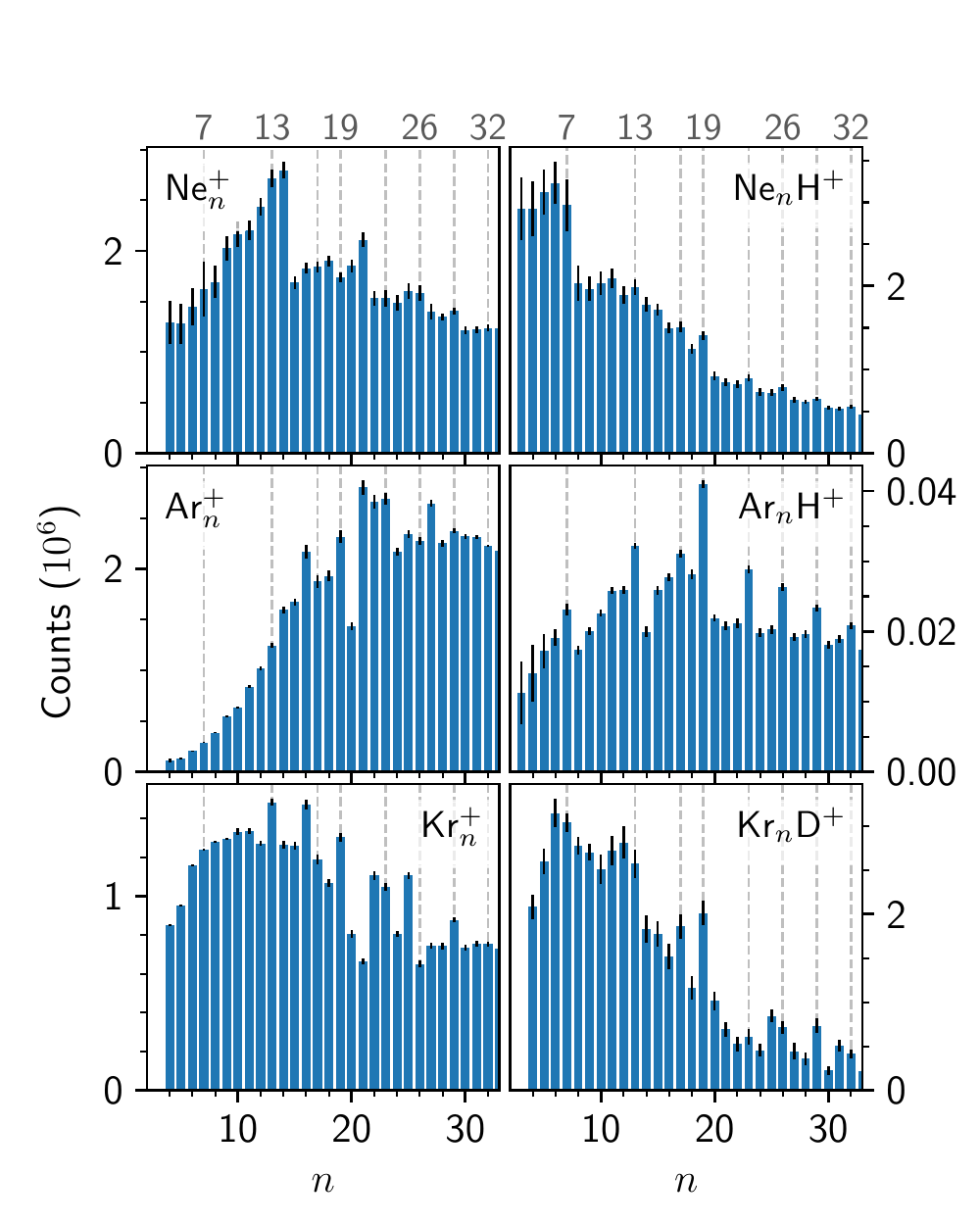}
\caption{Extracted intensites of Ne$_n^+$, Ne$_n$H$^+$, Kr$_n^+$, and Kr$_n$D$^+$ clusters from our measurements. Values for Ar$_n^+$ and Ar$_n$H$^+$ are from ref.\ \cite{Gatchell:2018aa}. Statistical errors from the measurements and fitted cluster distributions are indicated by the black bars.}
\label{fig:dists}       % Give a unique label
\end{figure}

Compared to argon, both neon and krypton have richer natural isotopic distributions. Argon is nearly mono-isotopic, with $^{40}$Ar making up 99.6\% of the natural abundance. On the other hand, Ne has two isotopes with abundances greater than 1\% and Kr has five. The result of this is exemplified in the right panel of Figure \ref{fig:MS} where we show a zoom-in of a mass spectrum of krypton and deuterium clusters and the distribution of species containing 13 Kr atoms. Here we are able to identify more than 50 different peaks originating from these species, all separated by about 1\,u. The contributions from Kr$_n^+$, Kr$_n$D$^+$, and Kr$_n$D$_3^+$, determined using IsotopeFit \cite{Ralser:2015aa}, are shown in the same panel, highlighting the complexity in determining the relative abundances of the different species (the uncertainties of these fits are shown in Fig.\ \ref{fig:dists}). The increasing complexity with cluster size is the limiting factor in what sizes we can study. For $n \gtrsim 30$, we are unable to reliably discern between the contributions of Kr$_n$ clusters with different numbers of D atoms attached. A similar limitation is also encountered for Ne$_n$H$^+$.

In Figure \ref{fig:dists} we show the extracted abundances of cationic and protonated Ne and Kr clusters together with the values for Ar from ref.\ \cite{Gatchell:2018aa}, for cluster sizes up to 32. The vertical dashed line and labels at the top of the figure indicate the expected magic numbers associated with sphere packing \cite{Echt:1981aa,Wales:1997aa}.  In each case there is an underlying size distribution that results from the pickup statistics of the rare gas atoms by the neutral He droplets. On top of these, abundance anomalies are visible indicating particularly stable or unstable structures. For Ne$_n^+$, $n=14$ and 21 stand out as being particularly abundant and these values are also observed for He$_n^+$ clusters \cite{Schobel:2011aa}. However, for the protonated Ne$_n$H$^+$ clusters, different magic numbers appear. Here $n=7$ is particularly strong relative to $n=8$, and weaker abundance enhancements are found at values in agreement with the dashed lines. For Kr$_n^+$, $n=13$, 19, 23, and 29, values expected for the packing of spheres, stand out, as do a few others such as 16 and 25. For the deuteronated Kr$_n$D$^+$ clusters, there are some noticeable differences; $n=6$ and 7 are particularly abundant, as is 17. Here, $n=13$ and 19 appear to be magic too, as for the pure clusters.

\begin{figure}
  \includegraphics[width=3in]{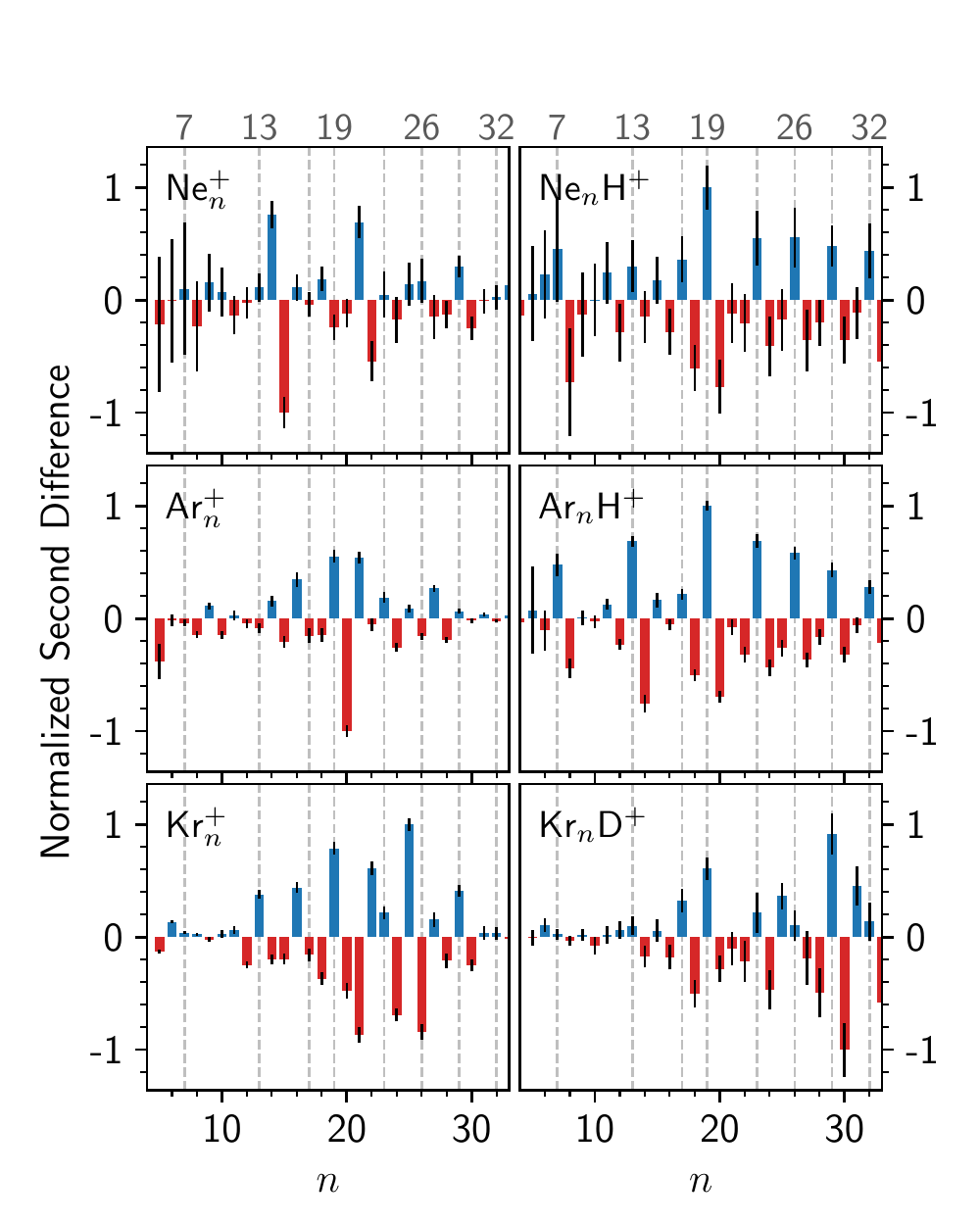}
\caption{Second differences calculated from the measured cluster intensities. The values are normalized to the largest absolute value and the statistical uncertainties are given by the black bars.}
\label{fig:diff}       % Give a unique label
\end{figure}

While some peaks stand out when comparing the size distributions in Figure \ref{fig:dists}, the underlying distributions make the different panels difficult to compare directly. In Figure \ref{fig:diff} we instead show the second differences of the cluster intensities, defined as

\begin{equation}
\Delta_2 = \ln\left[\frac{I_n }{\frac{1}{2}(I_{n-1} + I_{n+1})}\right],
\end{equation}
 where $I_n$ is the measured counts for cluster size $n$. Here the second difference values have also been normalized to the largest (absolute) value. Positive values $\Delta_2$ indicate cluster sizes that are more abundant than the mean of their neighbors ($n\pm1$). Likewise, negative values indicate clusters sizes with particularly low abundances compared to their neighbors.

\begin{figure}
\centering
  \includegraphics[width=1.75in]{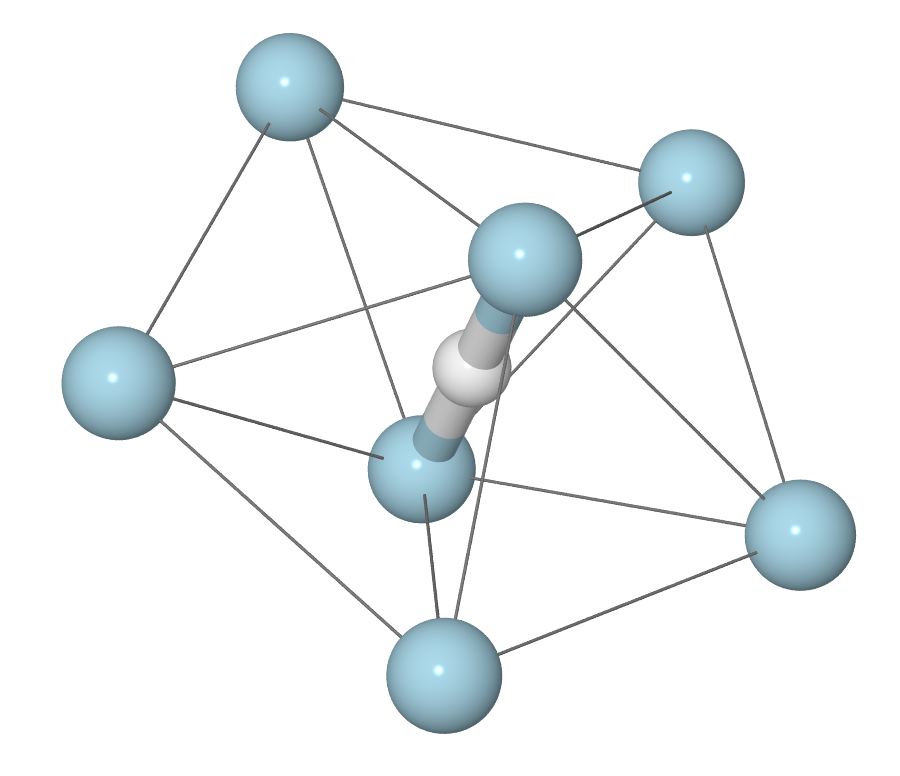}
\caption{Structure of Ne$_7$H$^+$ calculated at MP2(Full)/def2-SVPP level of theory. This structure is common to all protonated rare gas clusters and it is a magic size for each. The central Ne-H$^+$-Ne core is surrounded by a pentagon of Ne atoms, with their common plane passing through the proton. Thin connecting lines are used to highlight the external structure of the cluster. The coordinates for this structure are given in the electronic supplementary material.}
\label{fig:n7}       % Give a unique label
\end{figure}

From Figure \ref{fig:diff}, the magic sizes for each cluster series and the similarities and differences between the different systems become more obvious. For the Ne$_n^+$ and Ar$_n^+$, the $n=13$ magic number is suppressed with 14 being more abundant. Only for the protonated Ne and Ar clusters is 13 magic. For Kr on the other hand, 13 is indeed magic even for the pure cationic clusters. A similar trend is seen for $n=19$, one of the more prominent magic numbers in many of the data sets. For pure Ne$_n^+$ clusters, this is not a magic number, but for the pure Ar$_n^+$, Kr$_n^+$, and all of the protonated systems this is a highly abundant size. Overall the three protonated systems share many magic numbers, in good agreement with the values expected for sphere packing models (dashed line), showing that the stabilizing effect of the proton is visible for other rare gases than just Ar. A distinctive feature of the protonated clusters, in particular with Ne and Ar, is also the magic $n=7$ size, which corresponds to the first complete solvation sub-shell around the X-H$^+$-X core, an example of which is shown in Figure \ref{fig:n7}.

\begin{figure}
\centering
  \includegraphics[width=3in]{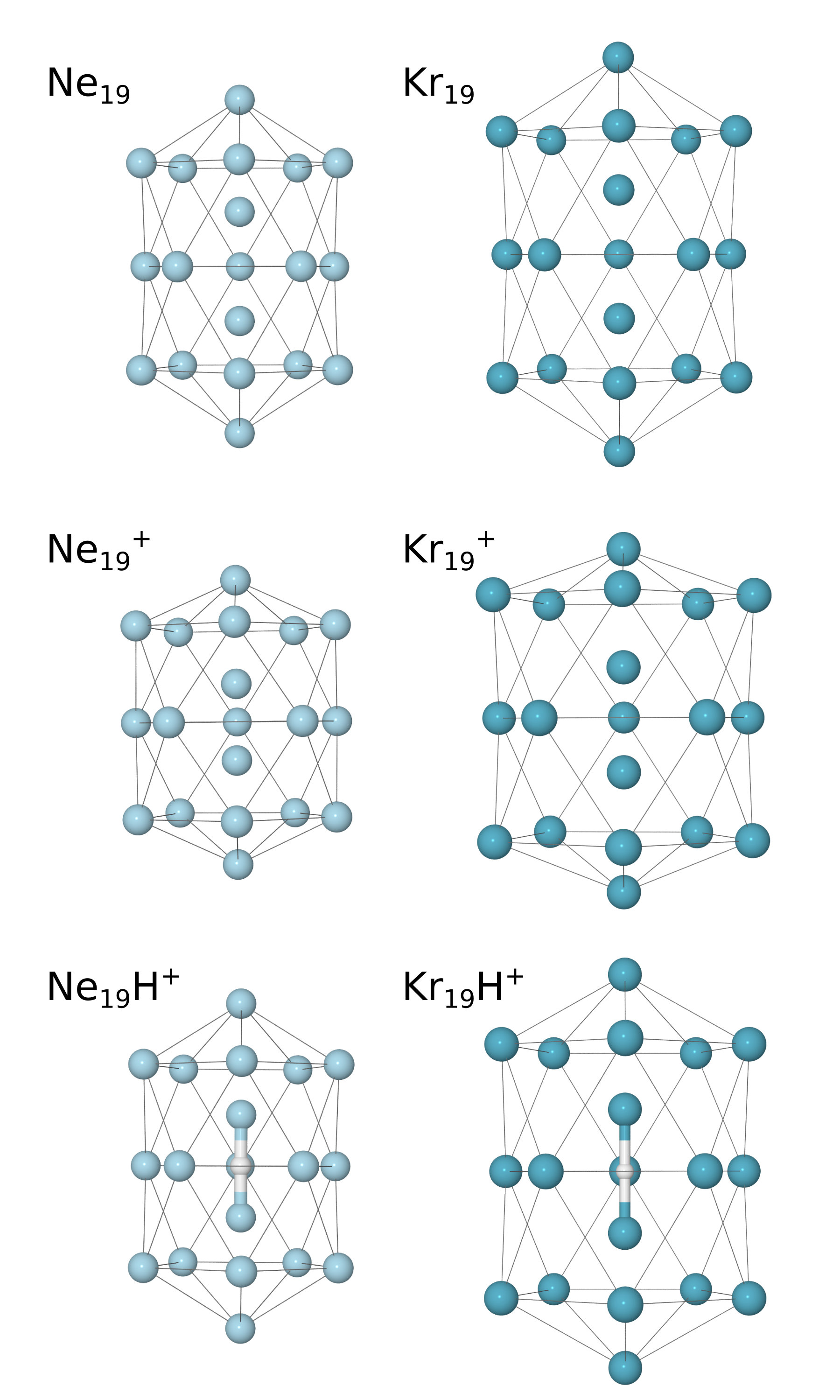}
\caption{Side views (perpendicular to symmetry axes) of the structures of neutral, cationic, and protonated clusters of neon and krypton containing 19 rare gas atoms calculated at MP2(Full)/def2-SVPP level of theory. Thin connecting lines are used to highlight the external structures of the clusters. The coordinates for these structures are given in the electronic supplementary material.}
\label{fig:n19}       % Give a unique label
\end{figure}
% For one-column wide figures uses

Compared to the lighter systems, the Kr$_n^+$ series shows a better agreement with the predicted magic numbers (dashed lines) and less of a difference between the pure and protonated systems. Most notably, $n=13$, 19, 25, and 29 are magic for both the pure and protonated clusters, and possibly other sizes too, like $n=23$. The reason for this is that the bond lengths of the charge center (e.g.\ Kr$_3^+$) is decreasingly contracted relative to the neutral atoms with increasing rare gas mass. This is seen in the comparison of the structures of neutral, cationic, and protonated clusters with 19 rare gas atoms shown in Figure \ref{fig:n19}. The neutral Ne$_{19}$ and Kr$_{19}$ are systems where all bond lengths are equal, giving highly symmetric (D$_{\text{5h}}$) structures. The removal of an electron changes the interactions in the clusters. In the case of neon, Ne$_{19}^+$ has a tightly bound covalent dimer as the charge center that causes significant distortion to the structure compared to the neutral system. The contracted core of the cluster could possibly allow two additional atoms to fit along the central axis of the cluster, which would explain the particularly hight abundance of Ne$_{21}^+$ in the experiments. In Kr$_{19}^+$, the charge is distributed over of a linear tetramer that, while having shorter bond lengths than the other atoms in the cluster, better matches the structure of the neutral system compared to Ne$_n^+$ and Ar$_n^+$ \cite{Gatchell:2018aa,Giju:2002aa}. When the charge is instead introduced as a proton, the separation between rare gas atoms surrounding the charge is increased and the structures are more similar to the neutral geometries. A Mulliken charge analysis shows that about 2/3 of the charge is carried by the proton in the case of the Ne$_{19}$H$^+$ clusters and 1/2 in the case of the Kr$_{19}$H$^+$ clusters (with the remaining positive charge shared equally by the closest two rare gas atoms). These findings are consistent with previous studies of Ar clusters, which show a behavior somewhere between that of Ne and Kr \cite{Gatchell:2018aa,Giju:2002aa}.

%For the Kr data, the uncertainties at large masses increase significantly due to the complex isotopic pattern, but 

\section{Conclusions}

Comparing the mass spectra of different cationic rare gas clusters, it is clear that the pure clusters often do not give the magic numbers that are commonly associated with them \cite{Echt:1981aa,Ding:1983aa,Harris:1984aa,Scheier:1987aa,Levinger:1988aa,Miehle:1989ab}. For Ne and Ar \cite{Gatchell:2018aa}, only the protonated clusters gives these characteristic numbers since the addition of a proton reduces the strain on the cluster geometry compared to the pure cationic systems. Pure Kr$_n^+$ clusters on the other hand \emph{do} show a better agreement with the formation of shell closures consistent with icosahedral symmetry, as do the protonated (here deuteronated) clusters. This suggests that clusters of heavier rare gases, e.g.\ Kr$_n^+$ and Xe$_n^+$, are less likely to see a strong difference in behavior compared to protonated clusters of the same species. For clusters of lighter rare gases (Ne and Ar), more care is needed to ensure that the magic cluster geometries indeed belong to the pure clusters and not, for instance, from protonated species that can easily form in the presence of (even trace amounts of) residual water.

%\begin{table}
%\begin{tabular}{lll}
%\hline\noalign{\smallskip}
%\end{tabular}
%\end{table}

%\begin{acknowledgements}
%\end{acknowledgements}

 %\section*{Conflict of interest}

 %The authors declare that they have no conflict of interest.

 \section*{Acknowledgments}

This work was supported by the Austrian Science Fund FWF (projects P31149 and P30355) and the Swedish Research Council (Contract No.\ 2016-06625).

%\bibliography{/Users/Michael/Dropbox/Documents/Bibtex/Library.bib}   % name your BibTeX data base

\begin{thebibliography}{10}
\providecommand{\url}[1]{{#1}}
\providecommand{\urlprefix}{URL }
\expandafter\ifx\csname urlstyle\endcsname\relax
  \providecommand{\doi}[1]{DOI \discretionary{}{}{}#1}\else
  \providecommand{\doi}{DOI \discretionary{}{}{}\begingroup
  \urlstyle{rm}\Url}\fi

\bibitem{Echt:1981aa}
O.~Echt, K.~Sattler, E.~Recknagel, Physical Review Letters \textbf{47}(16),
  1121 (1981)

\bibitem{Ding:1983aa}
A.~Ding, J.~Hesslich, Chemical Physics Letters \textbf{94}(1), 54 (1983)

\bibitem{Harris:1984aa}
I.A. Harris, R.S. Kidwell, J.A. Northby, Physical Review Letters
  \textbf{53}(25), 2390 (1984)

\bibitem{Scheier:1987aa}
P.~Scheier, T.D. M{\"a}rk, International Journal of Mass Spectrometry and Ion
  Processes \textbf{76}(2), R11 (1987)

\bibitem{Levinger:1988aa}
N.E. Levinger, D.~Ray, M.L. Alexander, W.C. Lineberger, The Journal of Chemical
  Physics \textbf{89}(9), 5654 (1988)

\bibitem{Miehle:1989ab}
W.~Miehle, O.~Kandler, T.~Leisner, O.~Echt, The Journal of Chemical Physics
  \textbf{91}(10), 5940 (1989)

\bibitem{Gatchell:2018aa}
M.~Gatchell, P.~Martini, L.~Kranabetter, B.~Rasul, P.~Scheier, Physical Review
  A \textbf{98}(2), 022519 (2018)

\bibitem{Giju:2002aa}
K.T. Giju, S.~Roszak, J.~Leszczynski, The Journal of Chemical Physics
  \textbf{117}(10), 4803 (2002)

\bibitem{Wales:1997aa}
D.J. Wales, J.P.K. Doye, The Journal of Physical Chemistry A \textbf{101}(28),
  5111 (1997)

\bibitem{Ferreira-da-Silva:2009aa}
F.~Ferreira~da Silva, P.~Bartl, S.~Denifl, O.~Echt, T.D. M{\"a}rk, P.~Scheier,
  Physical Chemistry Chemical Physics \textbf{11}(42), 9791 (2009)

\bibitem{Harris:1986aa}
I.A. Harris, K.A. Norman, R.V. Mulkern, J.A. Northby, Chemical Physics Letters
  \textbf{130}(4), 316 (1986)

\bibitem{Milne:1967aa}
T.A. Milne, F.T. Greene, The Journal of Chemical Physics \textbf{47}(10), 4095
  (1967)

\bibitem{Schobel:2011aa}
H.~Sch{\"o}bel, P.~Bartl, C.~Leidlmair, S.~Denifl, O.~Echt, T.D. M{\"a}rk,
  P.~Scheier, The European Physical Journal D \textbf{63}(2), 209 (2011)

\bibitem{Kurzthaler:2016aa}
T.~Kurzthaler, B.~Rasul, M.~Kuhn, A.~Lindinger, P.~Scheier, A.M. Ellis, The
  Journal of Chemical Physics \textbf{145}(6), 064305 (2016)

\bibitem{Kuhn:2016aa}
M.~Kuhn, M.~Renzler, J.~Postler, S.~Ralser, S.~Spieler, M.~Simpson,
  H.~Linnartz, A.G.G.M. Tielens, J.~Cami, A.~Mauracher, Y.~Wang,
  M.~Alcam{\'\i}, F.~Mart{\'\i}n, M.K. Beyer, R.~Wester, A.~Lindinger,
  P.~Scheier, Nature Communications \textbf{7}, 13550 (2016)

\bibitem{Scheidemann:1993aa}
A.~Scheidemann, B.~Schilling, J.P. Toennies, The Journal of Physical Chemistry
  \textbf{97}(10), 2128 (1993)

\bibitem{Ellis:2007aa}
A.M. Ellis, S.~Yang, Phys. Rev. A \textbf{76}, 032714 (2007)

\bibitem{Mauracher:2018aa}
A.~Mauracher, O.~Echt, A.M. Ellis, S.~Yang, D.K. Bohme, J.~Postler, A.~Kaiser,
  S.~Denifl, P.~Scheier, Physics Reports \textbf{751}, 1 (2018)

\bibitem{Ralser:2015aa}
S.~Ralser, J.~Postler, M.~Harnisch, A.M. Ellis, P.~Scheier, International
  Journal of Mass Spectrometry \textbf{379}, 194 (2015)

\bibitem{Frisch:2016aa}
M.J. Frisch, G.W. Trucks, H.B. Schlegel, G.E. Scuseria, M.A. Robb, J.R.
  Cheeseman, G.~Scalmani, V.~Barone, G.A. Petersson, H.~Nakatsuji, X.~Li,
  M.~Caricato, A.V. Marenich, J.~Bloino, B.G. Janesko, R.~Gomperts,
  B.~Mennucci, H.P. Hratchian, J.V. Ortiz, A.F. Izmaylov, J.L. Sonnenberg,
  Williams, F.~Ding, F.~Lipparini, F.~Egidi, J.~Goings, B.~Peng, A.~Petrone,
  T.~Henderson, D.~Ranasinghe, V.G. Zakrzewski, J.~Gao, N.~Rega, G.~Zheng,
  W.~Liang, M.~Hada, M.~Ehara, K.~Toyota, R.~Fukuda, J.~Hasegawa, M.~Ishida,
  T.~Nakajima, Y.~Honda, O.~Kitao, H.~Nakai, T.~Vreven, K.~Throssell, J.A.
  Montgomery~Jr., J.E. Peralta, F.~Ogliaro, M.J. Bearpark, J.J. Heyd, E.N.
  Brothers, K.N. Kudin, V.N. Staroverov, T.A. Keith, R.~Kobayashi, J.~Normand,
  K.~Raghavachari, A.P. Rendell, J.C. Burant, S.S. Iyengar, J.~Tomasi,
  M.~Cossi, J.M. Millam, M.~Klene, C.~Adamo, R.~Cammi, J.W. Ochterski, R.L.
  Martin, K.~Morokuma, O.~Farkas, J.B. Foresman, D.J. Fox.
\newblock Gaussian 16 rev. a.03 (2016)

\end{thebibliography}

\end{document}